\documentstyle[aps,preprint]{revtex}
\tighten
\begin{document}
\draft
\title{\bf Searches for Skyrmions in the Limit of Zero g-Factor }

\author{$^{1}$D.R.~Leadley, $^{2}$R.J.~Nicholas, $^{3}$D.K.~Maude, $^{4}$A.N.~Utjuzh, 
$^{3}$J.C.~Portal, $^{5}$J.J.~Harris and $^{6}$C.T.~Foxon}

\address{$^{1}$ Department of Physics, University of Warwick, Coventry, CV4~7AL,~UK\\
$^{2}$Department of Physics, Clarendon Laboratory, Parks Road, Oxford, OX1~3PU,~UK\\
$^{3}$Grenoble High Magnetic Field Laboratory, MPI-CNRS, 25~Avenue des Martyrs BP~166, 
F-38042~Grenoble Cedex~9, France\\
$^{4}$Russian Academy of Sciences, Institute of High Pressure Physics, 142092~Troitsk, Moscow 
Region, Russia\\
$^{5}$Department of Electronic and Electrical Enginering, University College, London, 
WC1E~7JE,~UK\\
$^{6}$Department of Physics, Nottingham University, University Park, Nottingham, 
NG7~2RD,~UK}

\date{To appear in Semiconductor Science and Technology, 1998}

\maketitle

\newpage
\begin{abstract}

Energy gaps have been measured for the ferromagnetic quantum Hall effect states at $\nu=1$ and 3 
in GaAs/Ga$_{0.7}$Al$_{0.3}$As heterojunctions as a function of Zeeman energy, which is 
reduced to zero by applying hydrostatic pressures of up to 20~kbar. At large Zeeman energy the 
gaps are consistent with spin wave excitations. For a low density sample the gap at $\nu=1$ 
decreases with increasing pressure and reaches a minimum when the $g$-factor vanishes. At small 
Zeeman energy the excitation appears to consist of a large number of reversed spins and may be 
interpreted as a Skyrmion. The data also suggest Skyrmionic excitations take place at $\nu=3$. The 
width of the minimum at $\nu=1$ is found to decrease as the $g$-factor is reduced in a similar way 
for all samples.
\end{abstract}

\subsection{Introduction}

The magnetic fields at which the integer quantum Hall effect (IQHE) \cite{iqhe} occurs correspond 
to the formation of a strongly interacting electron gas which can have a number of novel and 
interesting excitations.  This is particularly true when an odd number of levels are occupied, which 
makes the system spin polarized.  At $\nu=1$ (where the filling factor $\nu=n_e\hbar/eB$ measures 
how many Landau levels (LL) are filled) the ground state should be regarded as a ferromagnet 
since all the spin down states in the lowest LL are occupied while all the spin up states are empty.  
In GaAs the single particle (SP) Zeeman energy (ZE) $g\mu_BB$ is very small $\sim0.3$~K/T, 
and Coulomb interactions are very significant.  This has led several authors to suggest that novel 
charged excitations with non-trivial spin order, known as charged spin-texture excitations or 
Skyrmions, may occur \cite{fertig,sondhi}. 

In this paper we report experiments that investigate the nature of the collective excitations in the 
region of vanishing Zeeman energy, where the Zeeman energy is controlled by the use of 
hydrostatic pressure. Initially we will explain what is meant by and examine the difference between 
spin waves and Skyrmions. The high pressure experiments will then be described and results 
presented for $\nu=1$ that under different conditions show excitation of spin waves and large 
Skyrmions. Attention will then turn to $\nu=3$ where the data also suggests Skyrmionic excitations 
at vanishing ZE.  

The IQHE occurs when the Fermi energy is in a mobility gap of the electronic density of states and 
at low temperature this leads to quantised plateaux in the off diagonal component of 
magnetoresistivity $\rho_{xy}$ and zeros in $\rho_{xx}$ at integer filling factors $\nu$. The 
temperature dependence of these resistivity components can be used to measure the size of the 
energy gaps $E_g$. For even integer $\nu$ the gaps correspond to the cyclotron energy 
$\hbar\omega_c$ arising from the orbital motion of electrons, which experiments correctly 
measured as 20~K/T in GaAs. At odd integers the single particle (SP) Zeeman energy is tiny 
compared to $\hbar\omega_c$, yet experimentaly the odd and even IQHE appear in a similar 
temperature regime. At finite temperature the depth of an IQHE minimum in $\rho_{xx}$ is 
determined by smearing of the Fermi function and it will be approximately 50\% developed when 
$kT\sim E_g/6$. Thus if the energy gap for odd IQHE were determined by a SP~ZE it would only 
be observable for $B/T>20$, requiring $T<50$~mK to observe the spin splitting below 1~T and 
80~T to see $\nu=1$ at 4.2~K. This is clearly at varience with transport experiments which always 
measure a much larger gap \cite{usher} and suggests that the excitations at odd integer $\nu$ are 
instead due to a collective motion within the two-dimensional electron gas (2DEG). The energy 
scale of these excitations can be expected to scale with the Coulomb energy \cite{kh}, and the 
resulting increased splitting is sometimes referred to as an exchange enhanced $g$-factor. 

The $\nu=1$ ferromagnetic ground state is significantly different from the more familiar 
Heisenberg ferromagnet since the spontaneous magnetization occurs in the presence of a quantising 
magnetic field, not at zero field, and the spins associated with the charge carriers are free to move, 
hence it is termed an `itinerant' ferromagnet \cite{fertig,sondhi}. Two types of charged excitations 
from the ferromagnetic ground state that produce a well separated spin up electron and spin down 
hole have been identified. The first is a spin wave (really a spin exciton) \cite{kh} whose energy 
depends on wavevector, usually given by the dimensionless quantity $kl_B$ where 
$l_B=\sqrt{\hbar/eB}$ is the magnetic length. At long wavelength, i.e.\ \mbox{$kl_B=0$}, 
corresponding to no spatial separation between the electron and the hole, the spin wave energy is 
equal to $g\mu_BB$ as measured in spin resonance experiments \cite{dobers}. Transport 
measurements are sensitive to the opposite limit where the electron and hole are well separated, i.e.\ 
large $kl_B$, and see the whole Coulomb exchange energy $E_c=e^2/4\pi\epsilon l_B$ which in 
GaAs is $50.55 \sqrt{B}$~K, much larger than the SP~ZE and closer to $\hbar\omega_c$. The 
energy to create a spin wave is $E_{sw}=g\mu_BB+\kappa E_c$, where $\kappa$ is the spin 
stiffness calculated to be $\sqrt{\pi/2}$ in the ideal case.

The second type of excitation is based on a spin texture that consists of a central reversed spin 
surrounded by rings of spin that gradually cant over until at the edge they are aligned with the 
external magnetic field. We will refer to such spin textures as Skyrmions, although strictly this term 
is reserved for objects of infinite extent at zero ZE. The essential differences between this two-
dimensional spin texture and a spin wave are that the net spin may be greater than one and on a 
path taken around the central spin there will be a change of spin orientation equivalent to a winding 
number of unity. In a system with zero ZE the Skyrmions should have infinite extent but for finite 
ZE they have a finite size that can be characterised by the number of reversed spins $R$ contained 
in the Skyrmion. When the filling factor moves away from $\nu=1$ the ground state will contain a 
number of Skyrmions (quasi-holes) or anti-Skyrmions (quasi-electrons) and this has been detected 
from the degree of spin ploarisation in nuclear magnetic resonance \cite{barret} and 
photoluminescence experiments \cite{aifer}. Both of these measurements suggest that $R\sim7$. In 
transport measurements at exactly $\nu=1$ the excitations consist of well separated Skyrmion--
anti-Skyrmion pairs which, for infinite Skyrmions, only cost half the exchange energy required for 
a large spin exciton, but $R$ times the ZE. One way to think of this is that the spin texture of the 
Skyrmions dresses the spin exciton. Eventually, at large enough ZE, $R=1$ and Skyrmions are 
indistinguishable from the undressed spin excitons. 

The balance between the SP~ZE and the Coulomb energy is determined by the parameter 
$\eta=g\mu_BB/E_c$ which determines whether Skyrmions with $R>1$ (small $|\eta|$) or spin 
waves (large $|\eta|$) will be the lowest lying excitations \cite{eta}. The crossover is calculated to 
be at $|\eta|=0.054$ \cite{fertig}. It should be noted that $\eta\propto \sqrt{B}$ so Skyrmionic 
excitations are expected to be favoured at low magnetic fields and small $g$-factors. To date two 
transport measurements have inferred the existence of Skyrmions. Increasing $|\eta|$ by tilting the 
magnetic field suggested a 7 spin excitation for $|\eta|\sim0.01$ \cite{eise}. In a narrow quantum 
well where $g$ is already reduced by penetration of the wavefunction into the AlGaAs barrier, it 
was decreased further by hydrostatic pressure becoming zero at 4.8~kbar where the energy gap at 
$\nu=1$ showed a minimum \cite{duncan}. This indicated a much larger Skyrmionic excitation 
consistent with $R=33$ when $|\eta|<0.002$.

The ground states at higher odd filling factors will also be ferromagnetic but the effect on transport 
measurments is expected to be less pronounced as only a fraction $1/\nu$ of the electrons are 
involved in the collective motion. The remainder in full LLs do not contribute to the transport 
current, but may act to screen any Coulomb interactions. Skyrmions also appear in the excitation 
spectrum at higher odd filling factors but for an ideal 2DEG they are calculated to have higher 
energies than the single spin exciton at vanishing ZE \cite{wu}. However, when the finite thickness 
$z$ of a real 2DEG is taken into account Skyrmions may become the lowest energy excitation at 
$\nu=3$ \cite{fertig2} and also at all other odd filling factors for sufficiently extended 
wavefunctions \cite{cooper}. The stability and size of the Skyrmions is predicted to increase with 
$z$ but to be reduced by finite ZE and LL mixing. For example, at $\eta=0$ Skyrmions become the 
lowest excitation at $\nu=3$ once $z>0.1l_B$, while for $z=l_B$ the transition occurs at 
$|\eta|=0.0037$. So although Skyrmions were not observed in Ref.~\cite{eise} with $|\eta|>0.007$, 
or in Ref.~\cite{dunc2}, where the sample mobilities were relatively low and measurements were 
made at high fields , this does not preclude their existence under more favourable conditions.

\subsection{High pressure experiments}

This paper reports experiments in which Skyrmion formation is favoured by reducing the $g$-
factor. This is achieved by applying hydrostatic pressures of up to 22~kbar \cite{pres}. In GaAs at 
ambient pressure $g=-0.44$, as a result of subtracting band structure effects driven by the spin-orbit 
interaction from the free electron value of 2. At higher pressure the band structure contribution 
reduces, and so does the magnitude of $g$ which passes through zero at $\sim 18$~kbar. The 
pressure for this zero crossing decreases slightly at higher magnetic field, as the cyclotron energy 
increases the energy separation between the electron and hole bands. The $g$-factor has been 
calculated using {\bf k.p} theory \cite{sst32,hermann} and may be approximated by the following 
expression:
\begin{equation}
g=2-19300\left(\frac{1}{1519+\hbar\omega_c+10.7P}-
\frac{1}{1860+\hbar\omega_c+10.7P}\right)-0.12,
\label{eqn:pres}
\end{equation}
where $P$ is the pressure in kbar, and all far band terms have been assumed to stay constant 
\cite{hermann}. 

The samples studied were high quality GaAs/Ga$_{0.7}$Al$_{0.3}$As heterojunctions grown by 
molecular beam epitaxy at Philips Research Laboratories, Redhill. Samples G586, G627 and G902 
have undoped spacer layers of 40, 40 and 20~nm. At ambient pressure and 4~K their respective 
electron densities $n_e$ after photoexcitation are 3.3, 3.5 and $5.7\times 10^{15}{\rm m}^{-2}$ 
with corresponding mobilities of 300, 370 and 200~m$^2$/Vs. The samples were mounted inside a 
non-magnetic beryllium copper clamp cell \cite{cell} and the pressure was measured from the 
resistance change of manganin wire. The absolute values quoted at low temperature are accurate to 
$\pm1$~kbar, but between data points the variation is less than $\pm0.2$~kbar. The pressure cell 
was attached to a top loading dilution refrigerator probe allowing temperatures as low as 30~mK to 
be obtained and measured with a ruthenium oxide resistor attached outside the pressure cell, which 
followed the sample temperature with a negligible time lag. Experiments were also performed at 
temperatures up to 15~K using a separate variable temperature cyrostat.

Increasing the pressure causes the GaAlAs conduction band to move relative to the GaAs 
conduction band in the well reducing the number of electrons. Above $\sim$13~kbar no electrons 
were present in the dark at low temperature, but a certain number could be recovered after 
illumination from a red LED. The illumination time required to obtain a constant number of 
electrons roughly doubled for every 2~kbar increase in pressure, reaching several hours at 20~kbar. 
The highest pressure studied was 22~kbar, but no conductivity could be measured despite 
prolonged illumination. The sample required several hours for the density to stabilize before 
quantitative measurements could be made during which it varied by less than 1\% over the full 
temperature range. The data from G586 was recorded with a density of $0.44\pm0.06\times 
10^{15}{\rm m}^{-2}$ above 13~kbar and slightly higher at lower pressures. This puts $\nu=1$ at 
1.8~T. For G627 and G902 the data was recorded over the wider density range 0.8--3.3$\times 
10^{15}{\rm m}^{-2}$ i.e.\ $\nu=1$ between 3~T and 14~T, but only for pressures up to 14~kbar 
using a Polish pressure cell and an InSb pressure guage. 

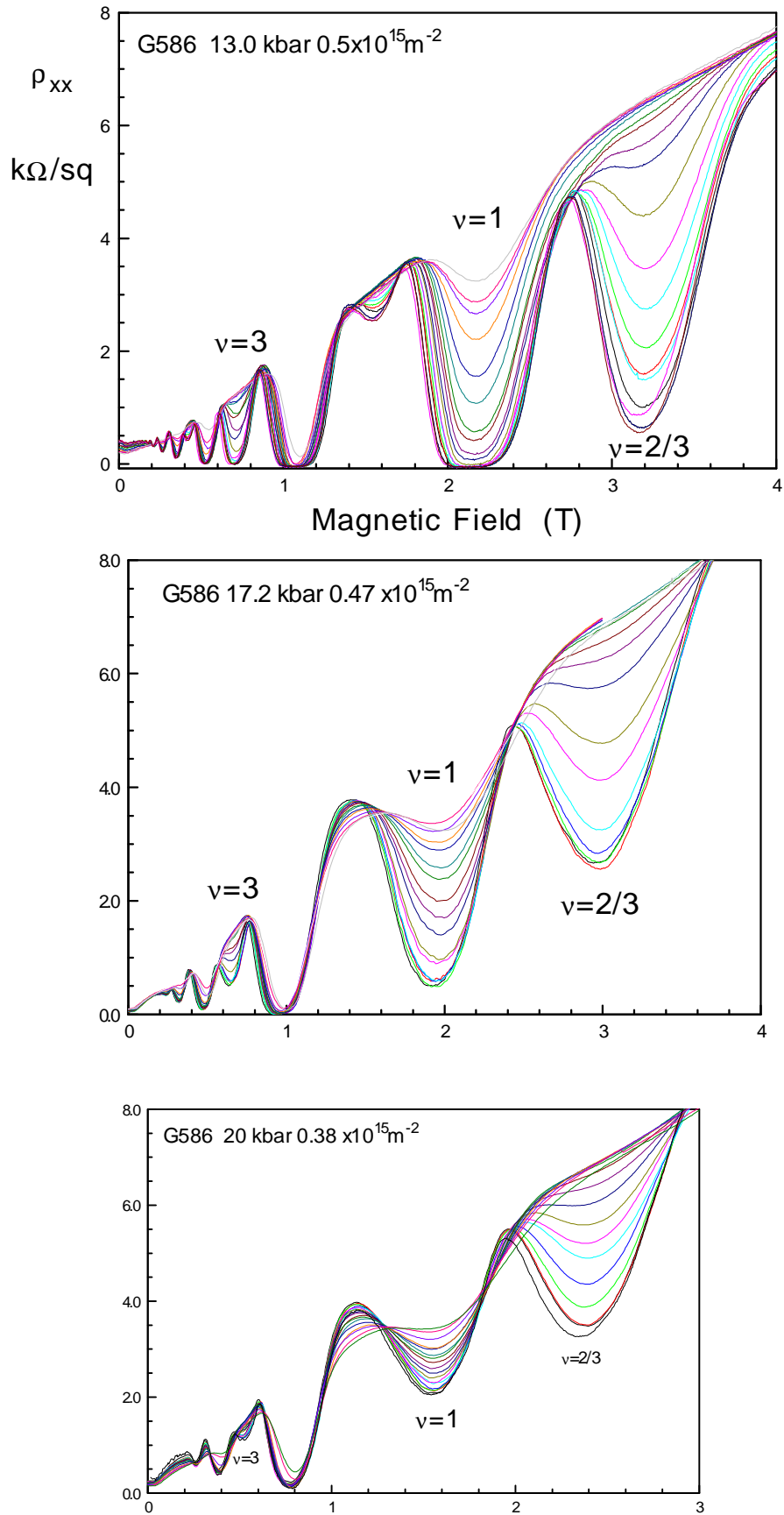
\begin{figure}
\caption{Magnetoresistance for sample G586 between 50~mK and 1.3~K at (a) 13~kbar, (b) 
17~kbar and (c) the highest pressure obtained of 20~kbar.}
\label{fig:xx586}
\end{figure}

Skyrmion formation is favoured by both lower magnetic fields and higher pressures, which reduce 
$|\eta|$.  In order to study this regime, data from sample G586 was taken for a constant 
$n_e\sim0.44\pm0.06\times 10^{15}{\rm m}^{-2}$, and up to higher pressures than the other two 
samples where $\nu=1$ is always at larger fields making the reduction of $|\eta|$ harder.  
Figure~\ref{fig:xx586} shows $\rho_{xx}$ at pressures of 13, 17 and 20~kbar for the temperature 
range 40--1300~mK. 

Values of the energy gaps have been extracted by fitting the temperature dependence of the 
resistivity minima to the Lifshitz-Kosevitch (LK) formula, which accounts for thermal smearing of 
the Fermi function. In this formula $\Delta\rho_{xx}\propto X/\sinh X$, where $X=2\pi^2kT/E_g$ 
and $\Delta\rho_{xx}$ is defined as $(\rho_{xx}(\infty)-\rho_{xx}(T))/\rho_{xx}(\infty)$, with 
$\rho_{xx}(\infty)$ the resistivity that would be observed in the absence of the IQHE. This 
procedure, described in more detail in Ref.~\cite{prb}, has the advantages over finding activation 
energies from an Arrhenius plot that, firstly, it measures the gap between LL centers not the 
mobility gap, and so is less sensitive to changes in disorder and secondly, an accurate zero of 
resistance is not required, which avoids any problems of parallel conduction and means especially 
low temperatures are not required. Examples of the fitted data are shown in Fig.~\ref{fig:lk}. A 
possible disadvantage of the LK method is that the majority of measurements are made in a 
temperature range in which the system may not remain totally spin polarized. The accuracy of the 
LK fitting procedure has been tested by considering the energy gaps at even integers which were 
found to be within 1\% of the expected $\hbar\omega_c$, e.g.\ in sample G586 at 10~kbar, with 
$\nu=1$ at 3.6~T, the gaps at $\nu=4$, 6, 8 and 10 were 15.7, 10.7, 7.9 and 6.4~K respectively. In 
general the odd $\nu$ data do not always fit the LK formula quite as well but we would expect the 
results to be accurate to $\pm$10\%.

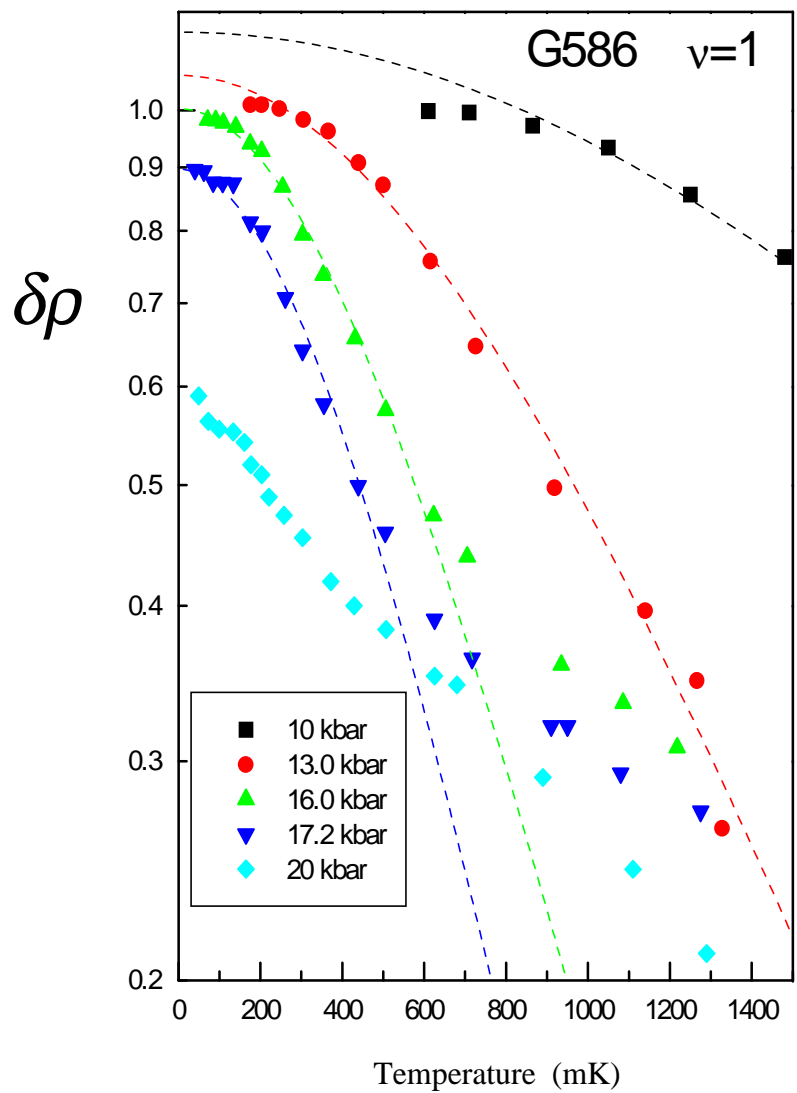
\begin{figure}
\caption{Arrhenius plot of the resistivity at $\nu=1$ for sample G586, including fittted lines 
from which the activation energy $\Delta$ is obtained.}
\label{fig:arrh}
\end{figure}

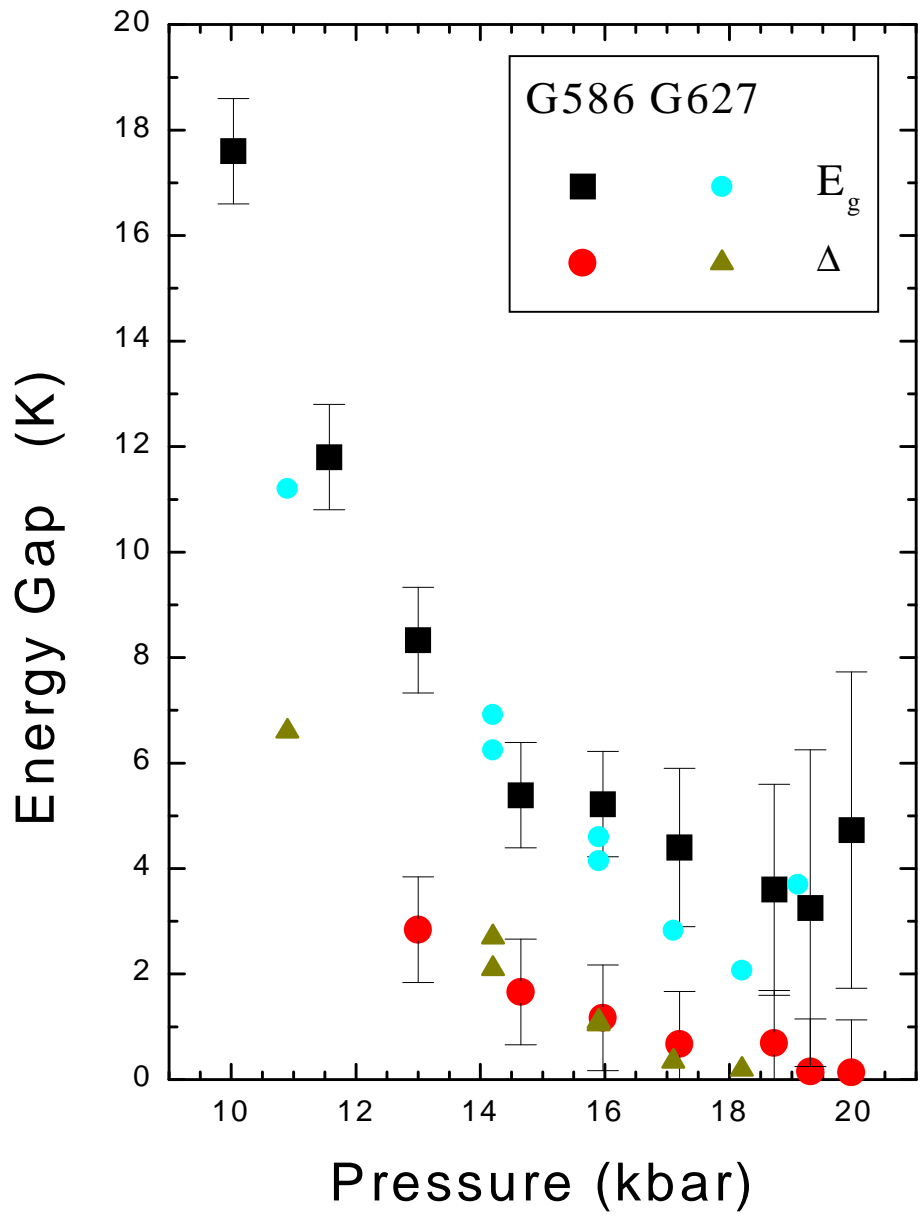
\begin{figure}
\caption{Energy gap at $\nu=1$ for sample G586 for pressures between 10 and 20~kbar. Notice 
how the gap decreases with pressure, until 18~kbar where $g=0$.}
\label{fig:eganddelta}
\end{figure}

We have also measured the activation energy $\Delta$ from an Arrhenius plot of 
$\rho_{xx}=\rho_0 \exp(-\Delta/2kT)$ as shown in Fig.~\ref{fig:arrh}. By contrast this only uses 
data at the lowest temperatures. A comparison of the results from the two techniques can be seen in 
Fig.~\ref{fig:eganddelta}, which shows the energy gaps $E_g$ deduced from the LK method and 
$\Delta$ from an Arrhenius plot in sample G586 for pressures in the critical range 10--20~kbar. 
The difference between the two values is due to the finite width $\Gamma$ of the extended states 
caused by Landau level broadening Provided the density remains unchanged this should be a 
constant such that $E_g = \Delta+\Gamma$. As the gap becomes small the LLs overlap, no well 
developed resistivity zero is observed and the activation behaviour collapses. This can be clearly 
seen for the 20~kbar data in Fig.~\ref{fig:arrh}. Consequently the values deduced from the 
Arrhenius plots become highly questionable.  Above 17~kbar even the LK fits fail systematically. 
This may be due in part to the fact that for the two highest pressures the maximum density achieved 
by prolonged illumination is significantly lower than for the lower pressures. At low temperatures 
the minima do not become zero so $\Delta\rho$ does not reach 100\%. Additionally at the high 
temperature end of our data range the resistivity shows an unusually slow temperature dependence, 
which would be interpreted as a very large energy gap if the LK formula were still valid. The 
values of the energy gaps shown on Fig.~\ref{fig:eganddelta} are for temperatures below this 
deviation from the $X/\sinh X$ law, but those at the highest pressures must still be regarded as 
relatively uncertain. Notwithstanding these qualifications the gap at $\nu=1$ clearly decreases as 
the pressure is increased. There is also some evidence from the higher temperature traces that it 
reaches a minimum at $\sim18$~kbar and beyond this pressure the gap recovers again, although the 
low temperature resistivity zero is not recovered. Before discussing the nature of the $\nu=1$ 
energy gap, we note that the existence of a symmetry about 18~kbar, however limited, is good 
evidence that the g-factor has really passed through zero at the pressure predicted by {\bf k.p}-
theory and indeed changed sign at the higher pressures. Further evidence of this can be seen at 
$\nu=1/3$, where the energy gap also collapses while the resistivity minimum dissappears and then 
reappears above 18~kbar \cite{cs}.

We do not believe that the collapse of the energy gap is due to pressure adversely affecting the 
mobility, for three reasons. First, at a given density the zero field mobility showed an initial 
increase but thereafter did not vary as the pressure was applied. Secondly, as can be seen in 
Fig.~\ref{fig:xx586} strong fractional QHE features were present at low temperature for all 
pressures \cite{sst32,cs} and in particular the feature at $\nu=2/3$ in sample G586 had an 
essentially constant energy gap for pressures between 10 and 20~kbar. Finally, the single particle 
lifetime $\tau_s$ has been obtained from a Dingle analysis of the low field, even integer, 
Shubnikov-de~Haas oscillations. Although there is some uncertainty in the values of $\tau_s$, 
since there are not many oscillations visible and the background resistivity is field dependent, the 
data above 13~kbar consistently yeild a value of $1.3\pm0.2$~ps, which suggests that the 
scattering of spin unpolarised electrons is not affected by the pressure.  Therefore we conclude that 
the change in energy gap at $\nu=1$ is connected with a change in spin stiffness caused by the 
effects of pressure on $g$.

\subsection{Spin waves at $\nu=1$}

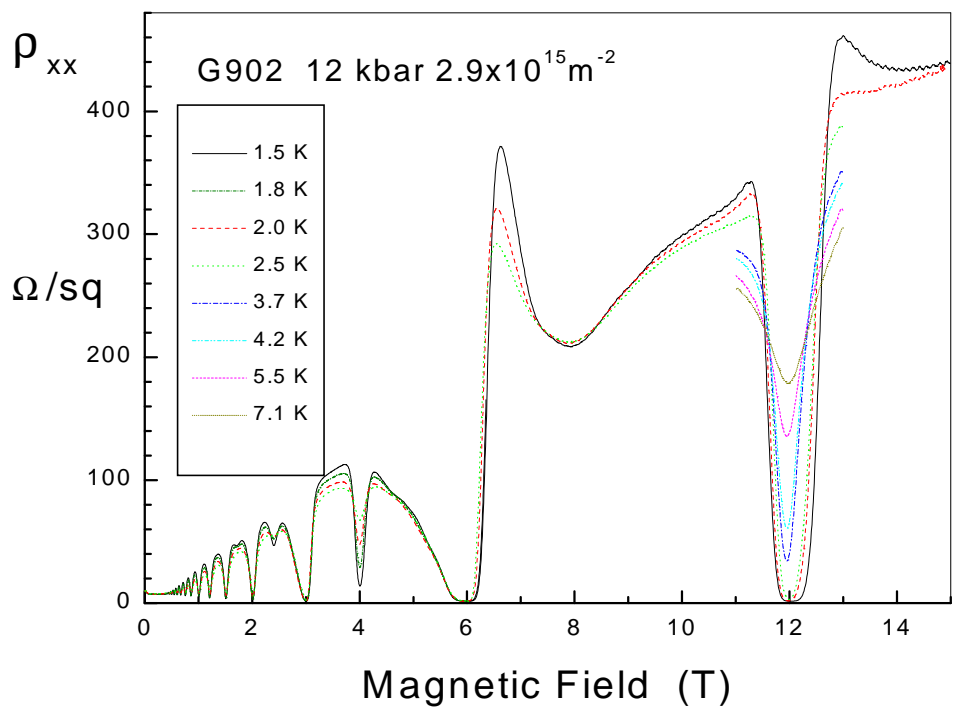
\begin{figure}
\caption{Energy gaps at odd integer $\nu$ for G627 and G902. The dashed line follows 
$n_e^{0.5}$ expected for spin wave excitation. Data for $\nu=3$ and 5 has been plotted at the 
effective density $n_e/\nu^2$.}
\label{fig:evn}
\end{figure}

Before deciding whether these observations provide evidence for Skyrmions we will consider data 
taken from the two higher density samples at relatively lower pressures, for which Skyrmion 
formation is less likely when $\nu=1$ occurs at high field. Typical magnetoresistance data from 
sample G902 is shown in Fig.~\ref{fig:xx902} for temperatures between 1.5 and 7~K and values of 
$E_g$ for $\nu=1$ in samples G627 and G902 are shown in Fig.~\ref{fig:evn} as a function of 
carrier density.  For these samples the resistivity at $\nu=1$ always becomes zero at sufficiently 
low temperature and the data fits the LK formula very well. Figure~\ref{fig:evn} demonstrates that 
the carrier density is a parameter that unifies the data from both samples even though many 
different pressures were used. The dashed curve is the best fit of $E_g\propto\sqrt{n_e}$ and since 
at $\nu=1$ $l_B\propto n_e^{-0.5}$ this shows that the gap is dominated by the Coulomb energy 
expected for spin wave excitation.  The equation of the line is $E_g=0.22~E_c$ which gives a spin 
stiffness considerably smaller than the theoretical estimate of $\sqrt{\pi/2}$ (=1.25) for an ideal 
2DEG.  Often such descrepencies between theory and experiment can be explained by the finite 
thickness of the real 2DEG softening the Coulomb interaction, but this usually only accounts for a 
factor of 2 \cite{das}.  Our values are generally in line with previous experiments \cite{usher} so 
the source of discrepancy remains an open question.  The square root dependence reported here 
might be thought to be at varience with the linear behaviour of Ref.~\cite{usher}, however the gaps 
reported by Usher {\em et al.}\ \cite{usher} were activation energies ($\Delta$) which do not 
include the LL broadening. Thus although the large gaps at high density agree with our data the 
smaller gaps are underestimated. The current data has also been analysed using Arrhenius plots and 
shows $\Delta$ increasing linearly with $n_e$ for each sample but with a sample dependent offset 
that increases with disorder. The universal curve of Fig.~\ref{fig:evn} can only be obtained when 
the LL broadening is correctly accounted for in the analysis. 

Also included in Fig.~\ref{fig:evn} are the energy gaps at $\nu=3$ and 5, which have been plotted 
at the equivalent density $n_e/\nu^2$. The squared filling factor is required to account both for 
lower magnetic field and the reduced number of electrons participating in the collective motion. It 
is seen that the universal curve for $\nu=1$ also describes the energy gaps at higher odd integer 
$\nu$ which suggests the gaps are all determined by the same mechanism of spin wave excitation. 
In fact, enhancement of the spin gap at higher odd integers appears always to give the same spin 
stiffness provided the correlation energy is greater than the disorder potential \cite{fogler,dlspin}.

\subsection{Skyrmions at $\nu=1$}
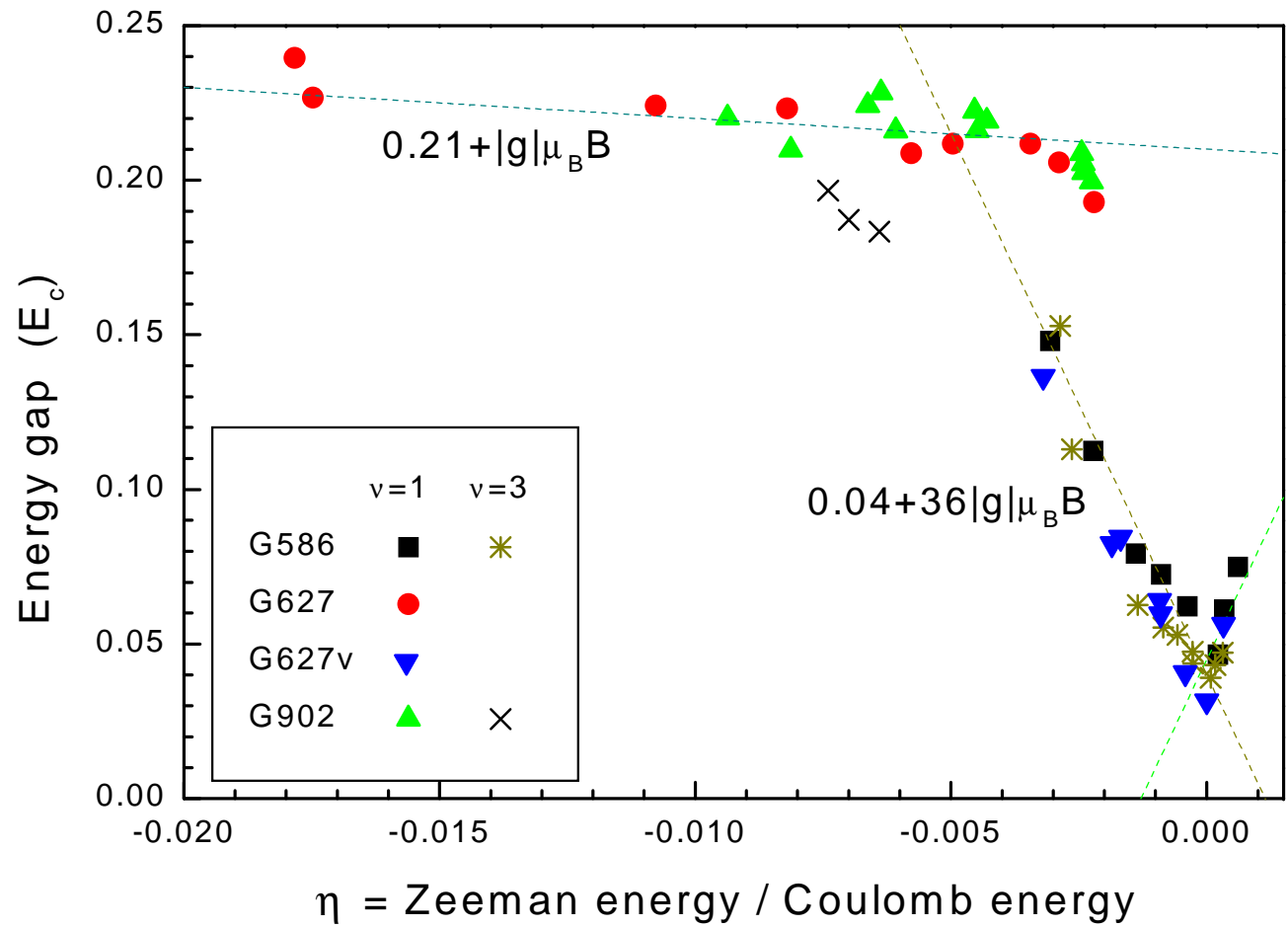
\begin{figure}
\caption{Energy gaps for all the samples as a function of ZE. Note that both axes are scaled by the 
Coulomb energy $E_c$. Solid points are for $\nu=1$, open for $\nu=3$. The dotted lines with 
gradients $\pm1$ show the energy to create a single spin exciton. The dashed lines have gradients 
of $\pm36$ corresponding to Skyrmion excitation.}
\label{fig:eta}
\end{figure}

We now return to the search for Skyrmions. In order to 
compare the experimental data with theory, the gaps have been scaled by the Coulomb energy and 
plotted as a function of $\eta$ in Fig.~\ref{fig:eta} for all the samples. On such a plot an energy gap 
that scales only with the Coulomb energy would show up as a horizontal line and a single particle 
spin gap would follow a line with unit gradient passing through the origin. For samples G902 and 
G627 the gap at $\nu=1$ appears to scale with the Coulomb energy plus the much smaller SP~ZE, 
represented by the dotted line with unit gradient. This corresponds to the simple spin-wave as 
expected from the above discussion. By contrast the data taken above 9~kbar for G586 with 
$|\eta|<0.0035$, exhibits a rapid change proportional to the ZE but with a slope much greater than 
unity. The dashed lines on Fig.~\ref{fig:eta} have gradients of $\pm$36 which describe the data 
well at small $|\eta|$. According to the arguments of Refs.~\cite{eise,duncan} this suggests the 
energy gap has a component $36g\mu_BB$ and indicates an excitation involving the reversal of 
thirty six spins.

Theory suggests that the Skyrmion size should increase continuously as $|\eta|$ is reduced so this 
value of $R=36$ should be taken as only an average or limiting value. Kamilla, Wu and Jain 
estimated the number of reversed spins in an anti-Skyrmion by minimising its energy \cite{kwj} 
$E(R)/E_c=0.313+0.23 exp(-0.25R^{0.85})+\eta R$. This shows an anti-Skyrmion with 18 
reversed spins (i.e.\ 36 in the pair excitation) would occur at $|\eta|=0.0017$ which falls right in the 
middle of our data range, and that $R$ falls to 11 by $|\eta|=0.0050$. However, the minimum 
energy of 0.313~$E_c$ at $\eta=0$, which corresponds to a pair gap of 0.627~$E_c$, is very much 
larger than the 0.04~$E_c$ observed experimentally.

Careful inspection of Fig.~\ref{fig:eta} suggests that the gap for the other samples may also be 
about to fall once $|\eta|<0.003$. If this were the case the same analysis would suggest Skyrmionic 
excitations of even larger numbers of spins.  However there is a large uncertainty in this number 
due to the uncertainty both in the absolute value of pressure and the precise pressure at which the 
ZE will be zero, which is slightly magnetic field dependent. This uncertainty is much less for G586 
due to the rapid decrease in gap which is observed, and the suggestion of a minimum energy gap 
which allowed us to confirm the pressure where $g=0$.

We are led to deduce that the excitations contain these very large numbers of spins because the gap 
does not change until $|\eta|$ is quite small and then drops to a very small value. The experiment 
suggests that the minimum gap for Skyrmionic excitations is $0.04 E_c$ compared to $0.21 E_c$ 
for the spin wave gap at vanishing ZE. This is substantially different from the prediction of exactly 
a 50\% reduction, made in Ref.~\cite{sondhi}, for infinite sized Skyrmions. However, since the 
experimental gap for creating a spin wave is already 5 times smaller than the theoretical prediction 
it is not surprising that quantitative agreement with the Skyrmion theory is incomplete. It may be 
that the more fundamental question to address is why the spin wave energy is so small or 
equivalently why real samples have such low spin stiffness. Another qualitative difference from the 
theory is that instead of the cusp which would result if $R\rightarrow\infty$ as $g\rightarrow 0$ the 
minimum of Fig.~\ref{fig:eta} is more rounded, as found in Ref~\cite{duncan}. This may be 
explained by long range disorder limiting the Skyrmion size. At the density of $0.44\times 
10^{15}{\rm m}^{-2}$ an 18 spin Skyrmion would have a radius of 1140~\AA\ which is already 
larger than the spacer layer thickness that usually determines the scale of the disorder potential in 
modulation doped structures. A theoretical estimate of how the Skyrmion size is limited in real 
systems would be very useful at this point.

\subsection{Width of the $\nu=1$ minima}

Transport data provides another measure of the LL structure through the width of the QHE plateaux 
in $\rho_{xy}$, or equivalently the minima in $\rho_{xx}$. This is a measure of the number of 
localised states that must be passed through before conduction can take place through the extended 
states. For a low mobility sample, with a large number of localised states, the plateaux become very 
wide at low temperature with extremely sharp risers and $\rho_{xx}$ consists of a set of $\delta$-
functions, i.e.\ $\delta\nu$, the width in filling factor of the $\nu=1$ minimum, is close to unity. In 
the limit of $T\rightarrow 0$ there will only be one extended state. By contrast high mobility 
samples which show a lot of structure between the integer plateaux and exhibit the FQHE would 
appear to have a large number of extended states at $T=0$. In that case the plateaux are very 
narrow and $\delta\nu\rightarrow 0$. 

Figure~\ref{fig:width627} shows how the widths of the minima at $\nu=1$ and 2 change with 
pressure for sample G627 at 40~mK and a constant density. The log scale focuses attention on the 
low resistivity region of $\rho_{xx}$ which shows when conduction through extended states 
begins. At $\nu=1$ $\delta\nu_1$ decreases dramatically as the pressure is increased. By contrast at 
$\nu=2$ $\delta\nu_2$ only changes by a very small amount which is related to the slight increase 
in mobility at higher pressure. It should also be noted that the minima are quite symmetrical about 
the integer filling factor showing that quasi-electrons and quasi-holes are localised to the same 
degree.

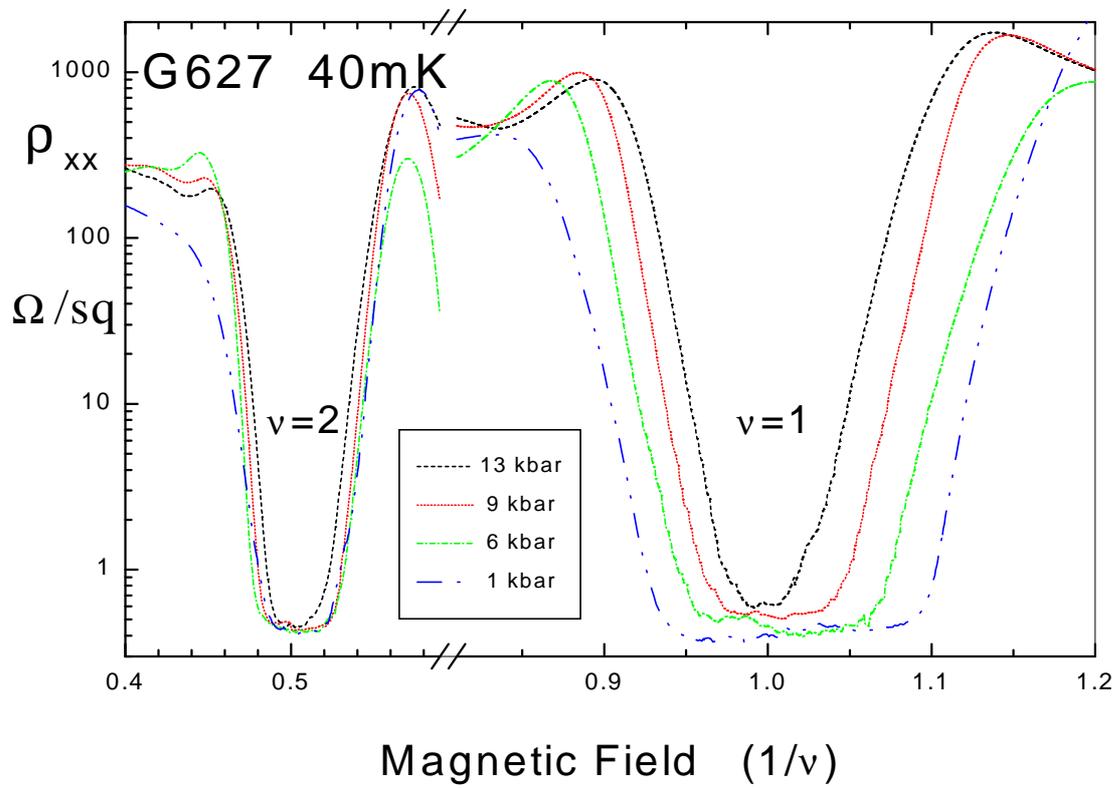
\begin{figure}
\caption{Ratio of the width of the minima at $\nu=1$ to $\nu=2$ showing a universal trend 
among all the samples to decrease with $\eta$. }
\label{fig:widths}
\end{figure}

We have measured $\delta\nu_1$ and $\delta\nu_2$ for each of the samples, at the lowest 
temperature possible, both at a fixed resistivity, in this case 10$\Omega/sq$, and at fixed fractions 
(1\% and 10\%) of $\rho_{xx}(\infty)$. As expected the plateaux widths vary significantly between 
samples, due to their different respective mobilities, and the criterion used to established the width. 
However, we find that the ratio $\delta\nu_1/\delta\nu_2$, with the same criteria used for each 
minimum, is much less sensitive to the temperature of the measurement or the actual value of 
resistivity chosen and shows a universal trend for all samples regardless of their mobility. This ratio 
is shown using the minima widths at 10$\Omega/sq$ in Fig.~\ref{fig:widths} as a function of 
$\eta$. (Data is not include for G586 at the highest pressures as the $\nu=1$ minima do not 
approach zero closely enough, i.e.\ the Hall plateaux are not flat, although they clearly exist at the 
correct value. However, if we look higher up in the minima, at the 50\% level, it appears that 
$\delta\nu_1/\delta\nu_2$ increases again above 18~kbar, i.e.\ the minima become wider again 
once $g$ has changed sign.) It is quite remarkable that the ratio $\delta\nu_1/\delta\nu_2$ appears 
to be independent of sample specific parameters like density or mobility, while the individual 
quantities $\delta\nu_1$ and $\delta\nu_2$ vary enormously. (We have also examined this ratio for 
a number of other structures with mobilities covering over two orders of magnitude, and at zero 
pressure this gives a ratio of $0.6\pm0.1$). As the ZE is varied most of the change in the ratio 
comes from $\delta\nu_1$ while $\delta\nu_2$ remains fairly steady for a given sample, although 
both widths vary with temperature in a similar way to make the ratio insensitve to temperature. If 
we interpret $\delta\nu_2$ as a measure of the number of localised single particle states then we 
might expect $\delta\nu_1/\delta\nu_2=1$ when conduction at $\nu=1$ is also by single particles. 
The ratio could be slightly less than unity due to the smaller gap at $\nu=1$ compared to the 
magnitude of the localisation potential and this would explain the experimental data at large 
$|\eta|$.  However, when collective phenomena are dominant at $\nu=1$, which we expect at small 
$|\eta|$, localisation will be very different.  For strong localisation we would expect a pinning of the 
states leading to a wider localised region, but only in the region close to $\nu=1$ where the 
Skyrmions are formed. This is the case for a Wigner crystal where it is only necessary to pin {\em 
one} particle to lock the whole lattice into place and make the system insulating. Thus strong 
localisation would predict that $\delta\nu_1$ should increase as $|\eta|$ decreases, contrary to the 
observations. For a weaker long range localising potential we might expect that a state will only be 
localised if {\em all} its constituent particles are localised. This means that for a fixed number of 
localisation sites there will be more extended states for collective motion than for single particle 
motion and so $\delta\nu_1$ will be less. A similar arguement would explain the observation that 
samples with narrow plateaux show the FQHE and vice versa. Thus Fig.~\ref{fig:widths}, where 
the ratio of the widths falls rapidly as $\eta$ and $g$ tend to zero, may be interpreted as showing 
how collective phenomena such as Skyrmions become more important at smaller $g$-factor. 

We still have to explain why $\delta\nu$ shows the same behaviour for all the samples, while 
$E_g$ did not. There is a significant difference between these two measurements. In the former 
case the filling factor is changed, i.e.\ flux quanta are added to or subtracted from the system which 
results in an excess of holes or electrons of reversed spin, which may be dressed to become 
Skyrmions or anti-Skyrmions. In the latter case a pair excitation is created with exactly one flux 
quantum per particle at $\nu=1$. It would thus appear that there is some difference between 
creating a single or paired excitation.

\subsection{Skyrmions at $\nu=3$}

We now turn to the data at $\nu=3$. As seen in Fig.~\ref{fig:evn} the energy gaps at $\nu=3$ for 
sample G902 are approximately consistent with the spin wave picture. When these data are scaled 
by $E_c$ at the relevent field and plotted on Fig.~\ref{fig:eta} it can be seen that they lie slightly 
below the $\nu=1$ data and decrease somewhat with $|\eta|$. The relationship between data from 
$\nu=1$ and 3 is different in the two figures since Fig.~\ref{fig:evn} considers the effect on the gap 
of the number of electrons participating whereas in Fig.~\ref{fig:eta} the parameter $\eta$ 
measures the ease of Skyrmion formation. Again there is a much larger effect seen in the results 
from G586. For this sample there is a dramatic decrease in gap between the two data points taken 
below 11~kbar and those at higher pressure. The pressure variation of the gaps deduced from the 
high pressure data is barely larger than their uncertainty, but it should be realised that the change in 
ZE is much smaller than over the same pressure range for $\nu=1$. When the data are correctly 
scaled and plotted on Fig.~\ref{fig:eta} they show a remarkable similarity to the data from 
$\nu=1$. This strongly suggests that the same mechanism is responsible for the excitations at both 
of these filling factors. Hence we are led to conclude that Skyrmionic excitations also occur at 
$\nu=3$ near to $g=0$.

Contrary to the ideal case \cite{wu}, this is indeed possible because the finite thickness $z$ of the 
2DEG has the effect of softening the Coulomb interaction once $z\sim l_B$ \cite{cooper}. For a 
density of 5$\times 10^{14}{\rm m}^{-2}$ the magnetic length at $\nu=3$ is 213~\AA\ and, using 
the variational method for a triangular well \cite{stern}, the mean distance of electrons from the 
interface is 219~\AA. According to Ref.~\cite{cooper} with $z=2 l_B$ Skyrmions will be excited 
at $\nu=3$ provided $|\eta|<0.0044$, which covers the whole range of the G586 data. It also 
predicts that at $|\eta|=0.002$, $R=15$ and $E_g=0.48E_c$. The first prediction is quite consistent 
with our data as a line drawn to include all the $\nu=3$ points on Fig.~\ref{fig:eta} would have a 
gradient of $\sim$22, but the size of the measured energy gaps is again an order of magnitude 
smaller than the theoretical predictions.

\subsection{R\^{o}le of disorder}

Finally it is necessary to consider whether this data really does provide evidence for large 
Skyrmions in the 2DEG of high mobility samples or if there is an alternative explanation for the 
precipitous drop in energy gap as $g=0$ is approached. It has been suggested that disorder may 
play an important r\^{o}le in determining the spin stiffness of the system \cite{fogler,green}. If the 
disorder potential is smaller than the ZE, it will play no significant role and spin waves will be 
created as normal. However, once the disorder potential is comparable to the ZE reversed spins will 
already exist in the ground state. This reduction in spin stiffness makes it is easier to perform 
additional spin flips so the spin waves become dressed, which in turn reduces the spin stiffness 
further. The excitations at very small ZE will thus contain many reversed spins. What is not clear is 
whether this mechanism will lead to a Skyrmionic spin texture or merely a multiple spin exciton. If 
the transition from single reversed spin excitations to multiple reversed spin excitations is critically 
driven by disorder then the energy gap may decrease more rapidly as $g=0$ is approached than was 
the case in the variable sized Skyrmion model discussed above, where there is a smooth change in 
size of the collective excitation. (We note that with the disorder driven picture it would not be 
possible to infer the numbers of spins involved in the excitations, either in this work or 
Refs.~\cite{eise} and \cite{duncan}.) So does this model of disorder induced Skyrmion formation 
fit with our observations? Figure~\ref{fig:eta} shows the drop in energy gap begins at larger 
$|\eta|$ for G586 than G627 or G902 but this is largely due to $\nu=1$ being at lower magnetic 
field. In fact the ZE where the drop begins is $\sim0.3$~K in each case, which would be expected 
as the samples are all fairly similar. However, this is a very small value for a disorder potential 
when compared with typical LL widths of several kelvin. Furthermore it is far from clear why the 
disorder could have such a dramatic effect on the exchange energy at $\nu=1$ without destroying 
the correlations responsible for the FQHE at $\nu=2/3$. It would have to be a strange type of 
disorder to affect the spin system without upsetting the spatial correlations. Finally, it is again hard 
to see how the minimum energy gap could be smaller than for an infinite sized Skyrmion--anti-
Skyrmion pair. Thus the disorder based explanation raises at least as many problems as it solves 
and at present there are no detailed theories with which to make comparisons. We therefore suggest 
that the data has a more convincing explanation in terms of Skyrmions at vanishing ZE.

Another possibility that we should consider is the phase separation at $g=0$ of a spin polarised 
$\nu=1$ system into an unpolarised system of two half filled Landau levels, similar to that 
observed in bilayers where the phases correspond to either $\nu=1$ in a single layer or $\nu=1/2$ 
in two layers. The decrease in gap and the strange temperature dependence of the highest pressure 
data might then be explained as a transition from a one component to two component phase. There 
are a number of reasons why we dismiss this possibility. First, we always observe a quantised Hall 
plateau which would not be expected for states based on $\nu=1/2$. Additionally in the bilayer 
system the two component phase is destroyed when the layer separation is small and for our 
situation there is no physical separation of the spin up and spin down electrons. The strong 
ferromagnetic interactions would then prevent any phase separation. However the possiblity 
remains of forming spatially separated spin up and spin down domains with the cost of forming 
domain boudaries being paid by the disorder potential. This would certainly limit the size of any 
Skyrmions and account for the rounded minimum in Fig.~\ref{fig:eta}, but causes problems in 
producing well separated Skyrmion--anti-Skyrmion pairs.

\subsection{Conclusion}

In summary we have measured the energy gaps for the ferromagnetic states at $\nu=1$ and 3 under 
conditions where the Zeeman energy can be tuned through zero. At large ZE the excitations involve 
a single reversed spin and are the well known spin waves. As the ZE is reduced to zero by applying 
hydrostatic pressure the energy gap decreases dramatically. At small ZE the excitations appear to 
consist of a large number of reversed spins which can be interpreted as Skyrmion-antiSkyrmion 
pairs. The same behaviour is seen both at $\nu=1$ and $\nu=3$ which suggests that the finite 
thickness of real 2DEGs makes Skyrmions the lowest lying excitations not only at $\nu=1$ but also 
at other odd filling factors.

\subsection*{Acknowledgements}

This work is supported by European Union TMR Programme number ERBFMGECT950077 
and NATO Research Grant 930471. DRL also recognises support from the ESPRIT programme 
through the Institute for Scientific Interchange Foundation.

\newpage

\begin{figure}
\caption{Temperature dependence of the $\nu=1$ minimum $\Delta\rho$ (as defined in the text) 
for sample G586 at pressures between 10 and 20~kbar. The dashed lines show fits to the LK 
formula. (NB only certain points are used in the fits as discussed in the text.)}
\label{fig:lk}
\end{figure}

\begin{figure}
\caption{Magnetoresistance for sample G902 at 12~kbar showing the temperature evolution of 
the minima at $\nu=1$ and 3.}
\label{fig:xx902}
\end{figure}

\begin{figure}
\caption{Resistivity as a function of inverse filling factor for sample G627 showing how the 
width of the $\nu=1$ minimum decreases with increasing pressure.}
\label{fig:width627}
\end{figure}

\begin{references}

\bibitem{iqhe} K.\ von Klitzing, G.\ Dorda and M.\ Pepper, Phys.\ Rev.\ Lett.\ {\bf 45}, 494 
(1980); T.\ Chakraborty \& P.\ Pietilainen, {\em The Quantum Hall Effects} (Springer-Verlag, 
New York, 1988); S.\ Das\ Sarma \& A.\ Pinczuk, {\em Perspectives in Quantum Hall Effects} 
(Wiley, New York, 1997) 

\bibitem{fertig} H.A.\ Fertig, L.\ Brey, R.\ Cote and A.H.\ MacDonald, \ Phys.\ Rev.\ B {\bf 50}, 
11018 (1994)

\bibitem{sondhi} S.L.Sondhi, A.Karlhede, S.A.Kivelson and E.H.Rezayi, Phys.\ Rev.\ B 
{\bf 47}, 16419 (1993)

\bibitem{usher} A.\ Usher, R.J.\ Nicholas, J.J.\ Harris and C.T.\ Foxon, Phys.\ Rev.\ B {\bf 41}, 
1129 (1990)

\bibitem{kh} C.~Kallin and B.I.~Halperin, Phys.~Rev.~B {\bf 30}, 5655 (1984)

\bibitem{dobers} M.\ Dobers, K.\ von~Klitzing and G.\ Weimann, Phys.\ Rev.\ B {\bf 38}, 
5453 (1988)

\bibitem{barret} S.E.\ Barret, G.\ Dabbagh, L.N.\ Pfeiffer, K.W.\ West and R.\ Tycko, Phys.\ Rev.\ 
Lett.\ {\bf 74}, 5112 (1995) 

\bibitem{aifer} E.H.\ Aifer, B.B.\ Goldberg and D.A.\ Broido, Phys.\ Rev.\ Lett.\ {\bf 76}, 680 
(1996)

\bibitem{eta} Some workers use the symbol $\tilde{g}$ instead of $\eta$ also note that in 
Ref.~4 $g$ is defined as half the Land\'{e} g-factor.

\bibitem{eise} A.\ Schmeller, J.P.\ Eisenstein, L.N.\ Pfeiffer and K.W.\ West, Phys.\ Rev.\ Lett.\ 
{\bf 75}, 4290 (1995)

\bibitem{dunc2} D.K.\ Maude, S.\ Marty, L.B.\ Rigal, M.\ Potemski, J.C.~Portal, Z.\ Wasilewki, 
M.~Henini, L.~Eaves, G.~Hill and M.A.\ Pate, Physica B {\bf 249-251}, 1 (1998)

\bibitem{duncan} D.K.\ Maude, M.~Potemski, J.C.~Portal, M.~Henini, L.~Eaves, G.~Hill and 
M.A.~Pate, Phys.\ Rev.\ Lett.\ {\bf 77}, 4604 (1996)

\bibitem{wu} X.-G.\ Wu and S.L.\ Sondhi, Phys.\ Rev.\ B {\bf 51}, 14725 (1995)

\bibitem{fertig2} H.A.\ Fertig, L.\ Brey, R.\ Cote, A.H.\ MacDonald, A.\ Karlhede and S.L.\ 
Sondhi, Phys.\ Rev.\ B {\bf 55}, 10671 (1997)

\bibitem{cooper} N.R.\ Cooper, Phys.\ Rev.\ B {\bf 55}, R1934 (1997)

\bibitem{pres} N.G.\ Morawicz, K.W.J.\ Barnham, A.\ Briggs, C.T.\ Foxon, J.J.\ Harris, S.P.\ 
Najda, J.C.\ Portal and M.L.\ Williams, Semicond.\ Sci.\ Technol.\ {\bf 8} 333 (1993); S.\ Holmes, 
D.K.\ Maude, M.L.\ Williams, J.J.\ Harris, J.C.\ Portal, K.W.J.\ Barnham and C.T.\ Foxon, 
Semicond.\ Sci.\ Technol.\ {\bf 9}, 1549 (1994)

\bibitem{sst32} R.J.\ Nicholas, D.R.\ Leadley, M.S.\ Daly, M.~van~der~Burgt, P.\ Gee, J.\ 
Singleton, D.K.\ Maude, J.C.\ Portal, J.J.~Harris and C.T.~Foxon, Semicond.Sci.Technol.\ {\bf 
11}, 1477 (1996)

\bibitem{hermann} C. Hermann and C. Weisbuch\ Phys.\ Rev.\ B {\bf 15}, 823 (1977)

\bibitem{cell} M.I.\ Eremets and A.N.\ Utjuzh, High-Press.\ Sci.\ Techn., {\bf 2}, 1597 (1993)

\bibitem{prb} D.R.\ Leadley, M.~van~der~Burgt, R.J.~Nicholas, J.J.\ Harris and C.T.\ Foxon, 
Phys.\ Rev.\ B, {\bf 53}, 2057 (1996)

\bibitem{cs} D.R.\ Leadley, R.J.~Nicholas, D.K.\ Maude, A.N.\ Utjuzh, J.C.\ Portal, J.J.\ Harris 
and C.T.\ Foxon, Phys.\ Rev.\ Lett.\ {\bf 79}, 4246 (1997)

\bibitem{das} F.C.\ Zhang and S.\ Das\ Sarma, Phys.\ Rev.\ B {\bf 33}, 2903 (1986); T.\ 
Chakraborty, P.\ Pietilainen and F.C.\ Zhang, Phys.\ Rev.\ Lett.\ {\bf 57}, 130 (1986)

\bibitem{fogler} M.M.\ Fogler and B.I.\ Shklovskii, Phys.\ Rev.\ B {\bf 52}, 17366 (1995)

\bibitem{dlspin} D.R.\ Leadley R.J.~Nicholas, J.J.~Harris and C.T.~Foxon, 
{\em submitted to} Phys.\ Rev.\ B (1998); LANL cond-mat/9805332

\bibitem{kwj} R.K.\ Kamilla, X.G.\ Wu and J.K.\ Jain, Solid\ State\ Commun.\ {\bf 99}, 289 
(1996)

\bibitem{stern} F.\ Stern, Phys.\ Rev.\ B {\bf 5}, 4891 (1972)

\bibitem{green} A.\ Green; LANL cond-mat/9801324.

\end{references}
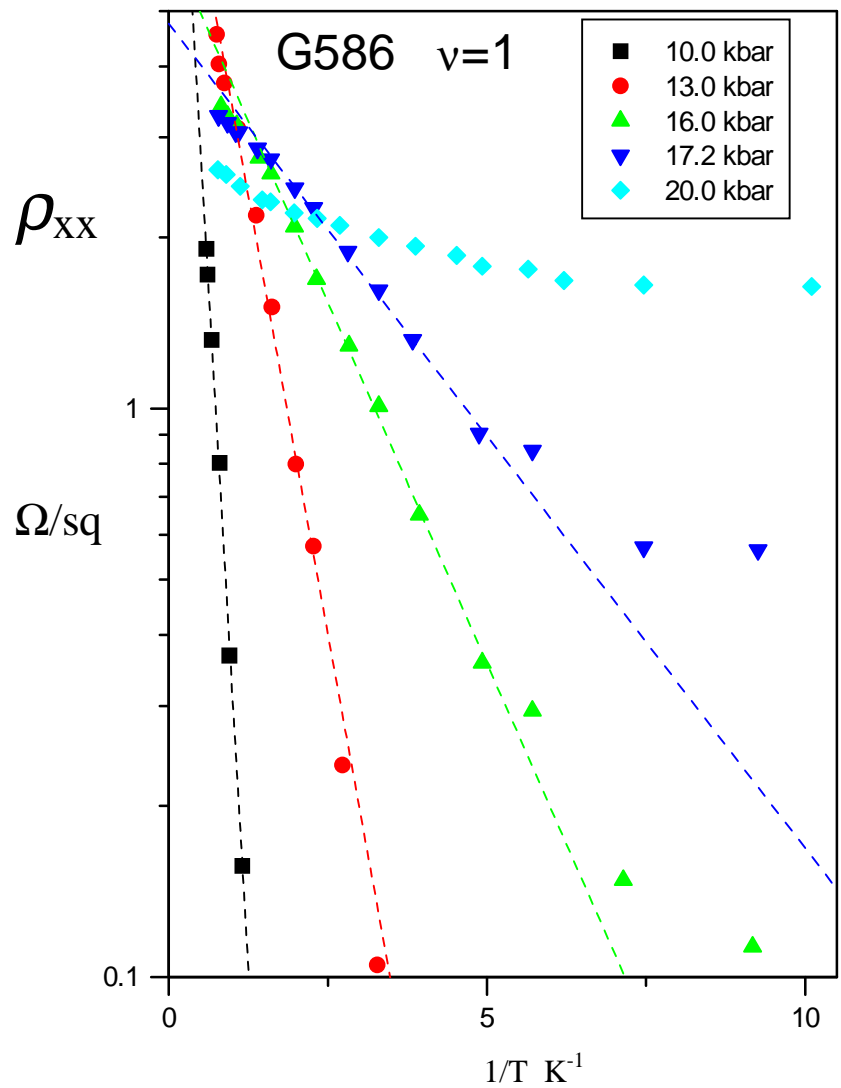
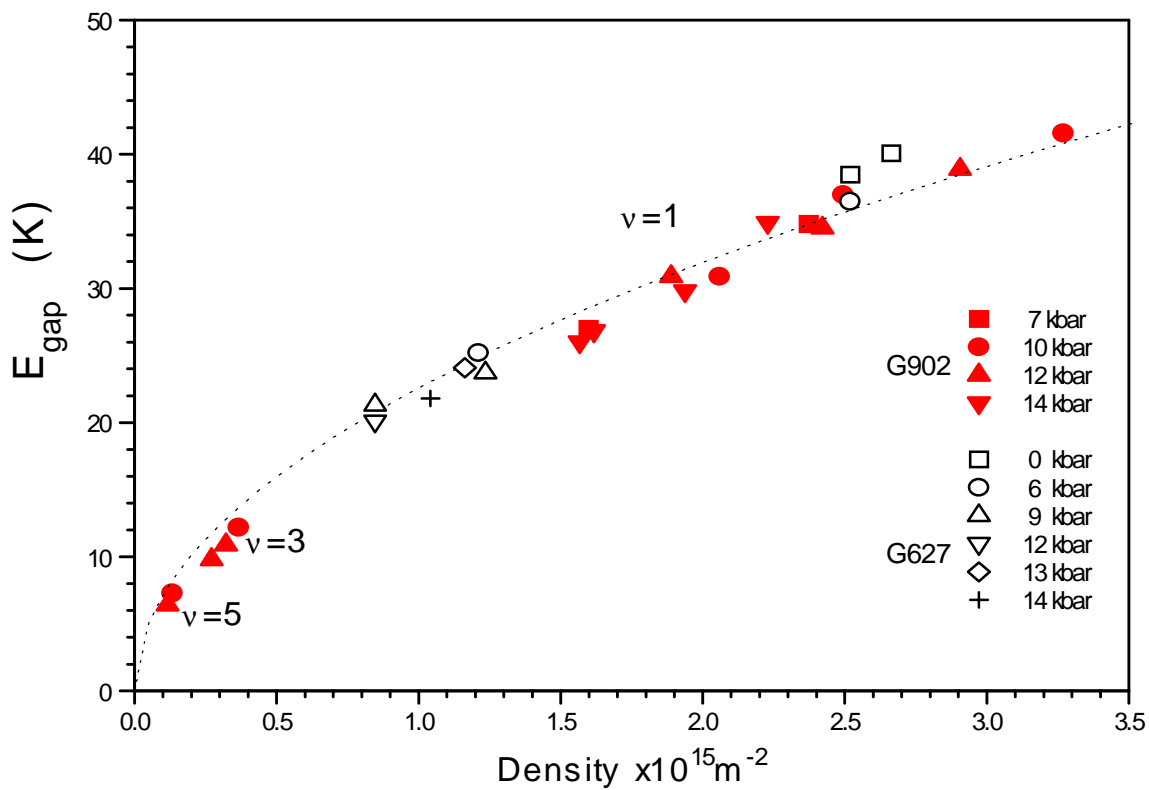
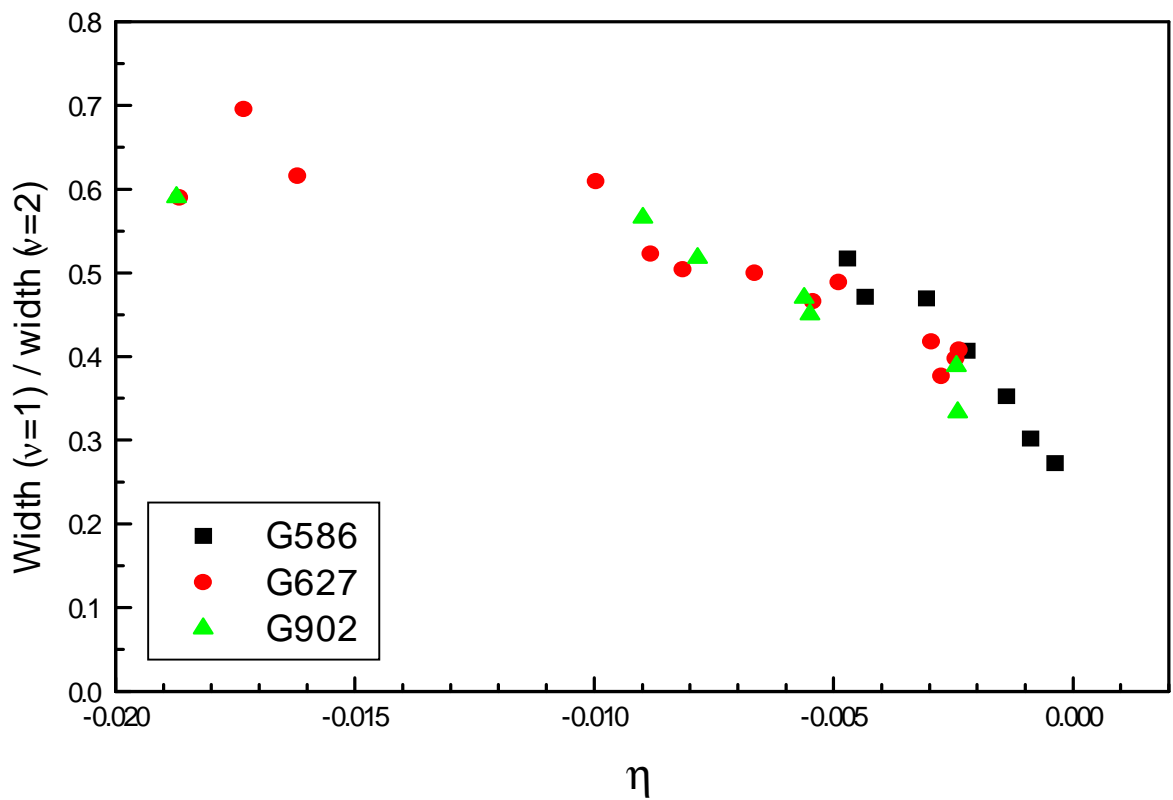
\end{document}